\documentclass[12pt]{article}
\usepackage{amssymb}

\usepackage{graphicx}
\usepackage{amsmath}
\usepackage{float}
\usepackage[dvips]{color}
\usepackage{rotating}

\input{tcilatex}

\begin{document}

\begin{flushright}
hep-th/0508109\\
\end{flushright}\vspace{1cm}

\begin{center}
{\Large \textbf{\ Soliton Solutions to the Einstein Equations in Five
Dimensions}}
\end{center}


\begin{center}
R. Clarkson\footnote{%
EMail: r2clarks@astro.uwaterloo.ca}, R. B. Mann\footnote{%
EMail: mann@avatar.uwaterloo.ca} \\[0pt]
Department of Physics, University of Waterloo, \\[0pt]
Waterloo, Ontario N2L 3G1, Canada\bigskip \\[0pt]
{\small We present a new class of solutions in odd dimensions to Einstein's
equations containing either a positive or negative cosmological constant.
These solutions resemble the even-dimensional Eguchi-Hanson--(anti)-de
Sitter ((A)dS) metrics, with the added feature of having Lorentzian
signatures. They provide an affirmative answer to the open question as to
whether or not there exist solutions with negative cosmological constant
that asymptotically approach AdS}$_{5}/\Gamma ${\small , but have less
energy than AdS}$_{5}/\Gamma ${\small . We present evidence that these
solutions are the lowest-energy states within their asymptotic class.}
\end{center}


Weakly coupled non-abelian gauge theories on non-simply connected manifolds
can have ground states of much lower energy than one might naively expect.
For example, a $U(N)$ gauge theory on a torus of $m$-dimensions whose
typical length is $L$ can have its lowest energy states of order $1/\left(
NL\right) $ (instead of $1/L$) if $N$ of the fields are arranged to be
periodic after traversing one circle $N$ times \cite{UNwraprefs}. By
introducing such locally flat but globally nontrivial connections, the
effective size of the compact space is thereby increased by a factor of $N$,
correspondingly reducing the spectrum of states. A string-theoretic
interpretation of this phenomenon is that of the low-energy excitations of
one D-brane wrapped $N$ times around a circle, where the $U(N)$ gauge theory
describes the low energy excitations of $N$ D-branes wrapped on the torus.
The $N^{2}$ open strings connecting distinct D-branes in the latter case
become $N$ multiply identified open strings on a circle of length $NL$ in
the former case, yielding a configuration with lower energy states.

The preceding construction can be generalized \cite{HoroJac} to quotient
spaces $M/\Gamma $, where $\Gamma $ is any freely acting discrete group of
isometries on the compact Riemannian manifold $M$, provided $N=n\left|
\Gamma \right| $, with $\left| \Gamma \right| $\ the number of elements in
the group. By wrapping $n$ branes $\left| \Gamma \right| $ times each around 
$M$, one obtains the same $U(N)$ gauge theory as if one had wrapped $N $
branes once each around $M/\Gamma $.

In this context the AdS/CFT correspondence conjecture implies the existence
of extra light states in a gauge theory formulated on a quotient space that
can be regarded as the boundary of an asymptotically AdS spacetime.
Specifically, the conjecture states that string theory on spacetimes that
asymptotically approach AdS$_{5}$ $\times $ $S^{5}$ is equivalent to a
conformal field theory (CFT) ($\mathcal{N}=4$ super Yang-Mills $U(N)$ gauge
theory) on its boundary $\left( S^{3}\times \mathbb{R}\right) \times $ $%
S^{5} $. Working at finite temperature (where the gauge theory is in a
thermal state described by the Schwarzschild-AdS solution) one can show \cite%
{HoroJac} that finite size effects on the gravity side become important at
high temperatures $T\backsim 1/\ell $ (where $\ell $ is the AdS radius). The
correspondence implies that the density of low energy states is not affected
even though the volume of $S^{3}$ has been reduced to $S^{3}/\Gamma $ --
hence there must exist light states of the type described above.

At zero temperature, taking the quotient by $\Gamma $ of AdS$_{5}$ produces
an orbifold with fixed points at $r=0$, and calculating the precise spectrum
of twisted sector states is difficult due to the Ramond-Ramond background.
However there are suggestive arguments \cite{HoroJac} that the AdS/CFT
correspondence still predicts the existence of extra light states. The
boundary energy of pure AdS$_{5}$ is exactly equal to the Casimir energy of $%
\mathcal{N}=4$ super $U(N)$ Yang-Mills on $S^{3}$ with radius $\ell $.
Taking the quotient by $\Gamma $ reduces both the energy and the volume by
the same factor, leaving the energy density unchanged.

The preceding conclusions would be modified if solutions of Einstein's
equations with negative cosmological constant existed that asymptotically
approached AdS$_{5}/\Gamma $ but had less energy. If so, then the ground
state energy density of the strongly coupled gauge theory on $S^{3}/\Gamma $
would be even smaller than on $S^{3}$.

The existence of such solutions has been an open question until now. Here we
show that there do exist solutions to Einstein's equations with cosmological
constant in 5 dimensions that are asymptotic to AdS$_{5}/Z_{p}$ where $p\geq
3$. They are obtained from a\thinspace generalization of the
Taub-Newman-Unti-Tamburino (Taub-NUT) metric \cite{MannStelea} in a manner
analogous to that used in deriving the AdS soliton. For large cosmological
constant their spatial sections approach that of the Eguchi-Hanson (EH)
metric \cite{EguchiHanson} and so we call these solutions Eguchi-Hanson
solitons. This is in contrast to the situation for the Kaluza-Klein
monopole, in which it has been shown that there are no static monopole
solutions to the five-dimensional Einstein equations with cosmological
constant that reduce to the asymptotically flat Kaluza-Klein monopole \cite%
{ontex}. The total energy of an EH soliton is negative, though bounded from
below, consistent with earlier arguments \cite{HoroJac}.

EH solitons are natural (5-dimensional) generalizations of the EH metric
that can be derived from a set of inhomogeneous Einstein metrics on sphere
bundles fibred over Einstein-Kahler spaces that were recently obtained \cite%
{MannStelea,LuPagePope}. Unlike the four dimensional case, a Lorentzian
signature is possible, thereby yielding a non-simply connected background
manifold for the CFT boundary theory.

To obtain the metric we begin with the five-dimensional generalization \cite%
{MannStelea} of the Taub-NUT metric, 
\begin{equation}
ds^{2}=-F(\rho )\left[ d\tau +2n\cos (\theta )d\phi \right] ^{2}+\frac{d\rho
^{2}}{F(\rho )}+(\rho ^{2}+n^{2})(d\theta ^{2}+\sin (\theta )^{2}d\phi
^{2})+\rho ^{2}dz^{2}  \label{5dTN}
\end{equation}%
where the $U(1)$-fibration is a partial fibration over a two-dimensional
subspace of the three dimensional base space. The function $F(\rho )$ is 
\begin{equation}
F(\rho )=\frac{4m\ell ^{2}-2n^{2}\rho ^{2}-\rho ^{4}}{\ell ^{2}(\rho
^{2}+n^{2})}  \label{f5dTN}
\end{equation}%
with $m$\ a constant of integration. The condition $n^{2}={\frac{\ell ^{2}}{4%
}}$ must hold for this to satisfy the 5D Einstein equations with positive
cosmological constant $\Lambda =\frac{6}{\ell ^{2}}$. Because of this the
metric (\ref{5dTN}) does not have a sensible $\Lambda \rightarrow 0$ limit.
In order to avoid singularities at $\theta =0,\pi $ the coordinate $\tau $
must be identified with period $8\pi n$, yielding a spacetime with closed
timelike curves.

We can obtain a solution with negative cosmological constant $\Lambda =-%
\frac{6}{\ell ^{2}}$ through a judicious choice of analytic continuations $%
z\rightarrow it,\tau \rightarrow 2n\psi $, $\ell \rightarrow i\ell $,
yielding the metric 
\begin{eqnarray}
ds^{2} &=&-\rho ^{2}dt^{2}+4n^{2}\tilde{F}(\rho )\left[ d\psi +\cos (\theta
)d\phi \right] ^{2}+\frac{d\rho ^{2}}{\tilde{F}(\rho )}+(\rho
^{2}-n^{2})d\Omega _{2}^{2}  \label{5dAdsTN} \\
\tilde{F}(\rho ) &=&\frac{\rho ^{4}+4m\ell ^{2}-2n^{2}\rho ^{2}}{\ell
^{2}(\rho ^{2}-n^{2})}  \label{ft5dTN}
\end{eqnarray}%
where $d\Omega _{2}^{2}$\ is the metric of the unit 2-sphere. By making the
further transformations 
\begin{equation}
\rho ^{2}=r^{2}+n^{2}\text{ \ \ \ \ \ }m=\frac{\ell ^{2}}{64}-\frac{a^{4}}{%
64\ell ^{2}}  \label{5dTNCHcoords}
\end{equation}%
and then setting $r\rightarrow r/2$, $t\rightarrow 2t/\ell $, we obtain 
\begin{eqnarray}
ds^{2} &=&-g(r)dt^{2}+\frac{r^{2}f(r)}{4}\left[ d\psi +\cos (\theta )d\phi %
\right] ^{2}+\frac{dr^{2}}{f(r)g(r)}+\frac{r^{2}}{4}d\Omega _{2}^{2}
\label{mtrcEHdS5} \\
g(r) &=&1+\frac{r^{2}}{\ell ^{2}}~~~,~~~~f(r)=1-\frac{a^{4}}{r^{4}}  \notag
\end{eqnarray}%
which solves Einstein's equations with negative cosmological constant $%
\Lambda =-6/\ell ^{2}$. Analytically continuing $\ell \rightarrow i\ell $
will turn (\ref{mtrcEHdS5}) into a metric solving Einstein's equations with
a positive cosmological constant.

The metric (\ref{mtrcEHdS5}) provides us with a new means of obtaining the
Eguchi-Hanson metric in 4-dimensions. In the $\ell \rightarrow \infty $
limit the metric (\ref{mtrcEHdS5}) yields the Eguchi-Hanson metric 
\begin{equation}
ds^{2}=\frac{r^{2}}{4}f(r)\left[ d\psi +\cos (\theta )d\phi \right] ^{2}+ 
\frac{dr^{2}}{f(r)}+\frac{r^{2}}{4} d\Omega_2^2  \label{4dEH}
\end{equation}
as a $t=$constant hypersurface. Note that the transformations (\ref%
{5dTNCHcoords}) are crucial in obtaining this result; the metric (\ref{5dTN}%
) becomes degenerate in the $\ell \rightarrow \infty $ limit.

The metric (\ref{mtrcEHdS5}) solves the Einstein equations with a negative
(positive) cosmological constant $\Lambda =\mp {\frac{6}{\ell ^{2}}}$ (or
the vaccum equations when $\ell \rightarrow \infty $). We call the metrics
with $\Lambda <0$ Eguchi-Hanson solitons, since they bear an interesting
resemblance to the EH metric in four dimensions and have a soliton-like
character similar to that of the AdS-soliton \cite{HoroMyers}. However
unlike the four dimensional case, a Lorentzian signature is possible; and
unlike the metric (\ref{5dTN}) there are no closed timelike curves (and no
horizons when $\Lambda <0$). Both the Ricci and the Kretschmann scalars are
easily seen to be free of singularities.

However string-like singularities can arise at $r=a$, and must be dealt with
separately. These can be eliminated in the usual way. Consider the the
behaviour of the metric (\ref{mtrcEHdS5}) as $r\rightarrow a$. Regularity in
the $\left( r,\psi \right) $ section implies that $\psi $ has period $\frac{
2\pi }{\sqrt{g(a)}}$ and elimination of string singularities at the north
and south poles $\left( \theta =0,\pi \right) $ implies that $\psi $ has
period $\frac{4\pi }{p}$ where $p$ is an integer. This implies in the
asymptotically AdS case that 
\begin{equation}
a^{2}=\ell ^{2}\left( \frac{p^{2}}{4}-1\right)  \label{EHmatch}
\end{equation}
with $p\geq 3$, yielding in turn that $a>\ell $.

We turn now to a computation of the energy of the EH soliton. Employing the
boundary-counterterm method \cite{AdSCFT,BaladS,Mann1,Mann2,GM1} we consider
the general gravitational action and add to it the counter-term action \cite%
{KLS}, which depends only on quantities intrinsic to the boundary and hence
leaves the equations of motion unchanged; it serves to cancel the
divergences of the usual bulk/boundary actions. In 5 dimensions it is 
\begin{eqnarray*}
I &=&-\frac{1}{16\pi G}\left[ \int_{\mathcal{M}}d^{5}x\sqrt{-g}\left(
R-2\Lambda \right) +2\int_{\partial \mathcal{M}}d^{4}x\sqrt{\gamma }\Theta
\right.  \\
&&\left. +\frac{2}{\ell }\int_{\partial \mathcal{M}}d^{4}x\sqrt{\gamma }%
\left( -3-\frac{\ell ^{2}}{2}\hat{R}\right) \right] 
\end{eqnarray*}%
where $\gamma $\ is the induced metric on the boundary whose
extrinsic/intrinsic curvature scalars are respectively $\Theta $ \ and $\hat{%
R}$. \ A conserved charge associated with a Killing vector $\xi $\ at
infinity can be calculated using the relationship 
\begin{equation}
\mathfrak{D}_{\xi }=\oint_{\Sigma }d^{d-1}S^{a}\xi ^{b}T_{ab}^{\text{eff}}
\end{equation}%
where\ $T_{ab}^{\text{eff}}$ is obtained by varying the full action
(including counter-terms, up to the appropriate dimension -- see refs. \cite%
{Mann1,KLS} for its explicit form) with respect to the induced boundary
metric, and $d^{d-1}S^{a}$ is the $(d-1)$ dimensional surface element
density.

For asymptotically AdS spacetimes, $\mathfrak{D}_{\xi }$ is the conserved
mass when $\xi ^{a}$ is a timelike-Killing vector; an analogous result holds
the asymptotically dS case \cite{GM1}. Thus, we find from the counter-term
method that the conserved mass (or total energy) is 
\begin{equation}
\mathfrak{M}=\frac{\pi (3\ell ^{4}-4a^{4})}{32G\ell ^{2}p}=\frac{(3\ell
^{4}-4a^{4})N^{2}}{16\ell ^{5}p}  \label{Mass}
\end{equation}%
where the second equality occurs because we can relate the parameters of the
gravity theory in the bulk to those of the CFT on the boundary, $G=\pi \ell
^{3}/(2N^{2})$. It is straightforward to show that this is equal to the
Euclidean action multiplied by the inverse of the (arbitrary) period of the
Euclidean time parameter, yielding a solution with zero entropy, as expected
for a horizonless metric.

We can compare the result (\ref{Mass}) with that of the field theory on the
boundary of the AdS$_{5}$ orbifold. Since the local geometry is unchanged
(with only the volume of the $S^{3}$ becoming that of $S^{3}/\Gamma $) the
calculation is the same as that for the AdS$_{5}$ case \cite{AwadJohnsonKerr}%
. Rescaling the AdS orbifold metric by a factor of $r^{2}/\ell ^{2}$, as $%
r\rightarrow \infty $ we find the metric of the conformal field theory 
\begin{equation}
ds^{2}=-dt^{2}+\frac{\ell ^{2}}{4}\left[ d\psi +\cos (\theta )d\phi \right]
^{2}+\frac{\ell ^{2}}{4}d\Omega _{2}^{2}  \label{mtrcCFT}
\end{equation}%
which has a vanishing Weyl tensor. The stress tensor is therefore that of a
field theory on a conformally flat spacetime in four dimensions whose
expectation value $\left\langle \hat{T}_{ab}^{s}\right\rangle $ is a known
quadratic function of the curvature with coefficients dependent upon the
field content of the theory \cite{BirrellDavies}. The energy is then given
by 
\begin{equation}
\mathcal{E}=\sum_{s=0,\frac{1}{2},1}n^{s}\int_{\Sigma }d^{3}x~\sqrt{\sigma }%
N_{lp}\left\langle \hat{T}_{ab}^{s}\right\rangle \xi ^{a}u^{b}
\label{EfromT4d}
\end{equation}%
where the sum is over the scalar, spinor and vector fields of the field
theory, and where $\theta \in \left[ 0,\pi \right] $, $\phi \in \left[
0,2\pi \right] $ and $\psi \in \left[ 0,4\pi /p\right] $. For $N=4$ Super
Yang-Mills theory there are 6 scalars, 4 spinors, and 1 vector field \cite%
{AwadJohnsonKerr}; inserting this information into (\ref{EfromT4d}) we
obtain 
\begin{equation}
\mathcal{E}=\frac{3N^{2}}{16\ell p}  \label{CasE}
\end{equation}%
Note that this matches the conserved mass given by (\ref{Mass}) with $a=0$.
A straightforward computation using a Noether charge approach \cite%
{GarfMann,Lorenzo} of the energy of the soliton (\ref{mtrcEHdS5}) relative
to the AdS orbifold yields the difference $\mathfrak{M}-\mathcal{E}$ as
expected. \ Note that going $p$ times along the $\psi $ direction the
situation is the same as if the asymptotic space were $S^{3}$ (and not $%
S^{3}/\Gamma $), for which fermions are periodic. Hence $p$ must be even
when fermions are present in the CFT.

We see that the energy (\ref{Mass}) of the EH soliton is lower than that of
the AdS$_{5}/\Gamma $\ orbifold, and is in fact negative and finite once the
condition (\ref{EHmatch}) is taken into account. Indeed, we have 
\begin{equation}
\mathcal{E}_{\text{EH-soliton}}=-\frac{(p^{4}-8p^{2}+4)N^{2}}{64\ell p}
\label{Energy}
\end{equation}
for any given integer $p\geq 3$. We conjecture that the EH soliton is the
state of lowest energy in its asymptotic class in both 5D Einstein gravity
with negative cosmological constant and in type IIB supergravity in 10
dimensions. Indeed, the AdS/CFT correspondence (along with the expected
stability of the gauge theory) suggests that any metric solving the 5D
Einstein equations that has the same boundary conditions as the EH soliton
will have a greater energy.

We now show that our conjecture holds perturbatively for all metrics of the
form 
\begin{equation*}
g_{\mu \nu }=\bar{g}_{\mu \nu }+h_{\mu \nu }
\end{equation*}%
where $\bar{g}_{\mu \nu }$\ is the EH solition (\ref{mtrcEHdS5}) and the
perturbation obeys the falloff conditions 
\begin{equation*}
h_{\mu \nu }=\mathcal{O}\left( r^{-2}\right) \text{ \ \ \ \ \ \ \ }h_{\mu
r}= \mathcal{O}\left( r^{-4}\right) \text{ \ \ \ \ \ \ \ }h_{rr}=\mathcal{O}
\left( r^{-6}\right) \text{ \ \ \ \ \ \ }\mu ,\nu \neq r
\end{equation*}
Employing the method of Abbot and Deser \cite{AbbotDeser}, the Hamiltonian $%
H $ on a time-symmetric slice to second order in the perturbation $h_{ij}$ ($%
i,j\neq t$) is 
\begin{equation}
\mathcal{H}=\bar{N}\left[ \frac{1}{\sqrt{\bar{g}}}p^{ij}p_{ij}+\sqrt{\bar{g}}
\left( \frac{1}{4}\left( \bar{D}_{k}h_{ij}\right) ^{2}+\frac{1}{2}{}^{(4)} 
\bar{R}^{ijkl}h_{il}h_{jk}-\frac{1}{2}{}^{(4)}\bar{R}^{ij}h_{ik}h_{j}^{%
\phantom{j}k}\right) \right]
\end{equation}
where we have used the same notation, gauge, and constraint equations as in
ref. \cite{HoroMyers}.

As the momenta make a positive contribution to the energy density, we need
only calculate the gradient energy density $\left( \bar{D}_{k}h_{ij}\right)
^{2}$ (also positive) and the potential energy density $U=\frac{1}{2}^{(4)} 
\bar{R}^{ijkl}h_{il}h_{jk}-\frac{1}{2}{}^{(4)}\bar{R}^{ij}h_{ik}h_{j}^{%
\phantom{j}k}$. We evaluate the latter by considering it as a matrix
contractied with two 9-vector whose components are $h_{il}$. We find that
there exists a negative eigenvalue for sufficiently small $r$. Writing the
peturbation as $h_{ik}=A\left( r\right) \tilde{h}_{ik}$, where $A\left(
r\right) $ is a profile function maximized at $r=a$\ and $\tilde{h}_{ik}$ is
the eigenvector associated with the negative eigenvalue, we find that the
negative potential energy $U$ is not outweighed by a simple estimation of
the gradient energy density (given by dividing the maximum of the profile
function by the proper distance over which $U$ is negative). This situation
-- quite unlike that for the AdS soliton \cite{HoroMyers} -- forces us take
into account the complete expression 
\begin{equation}
T = \left\{ \hat{h}_{ab} \partial_c A(r) + A(r) \hat{h}_{ab;c} \right\}^2
\end{equation}
for the gradient energy term. From this we find that the gradient energy
density always outweighs the potential energy density, indicating that the
EH soliton is perturbatively stable for all values of $p$ relative to all
other metrics with the same boundary conditions.

\bigskip We have also found $\left( d+1\right) $-dimensional generalizations
of the EH solition (\ref{mtrcEHdS5}). These are 
\begin{eqnarray}
ds^{2} &=&-g(r)dt^{2}+\left( \frac{2r}{d}\right) ^{2}f(r)\left[ d\psi
+\sum_{i=1}^{k}\cos (\theta _{i})d\phi _{i}\right] ^{2}  \notag \\
&&+\frac{dr^{2}}{g(r)f(r)}+\frac{r^{2}}{d}\sum_{i=1}^{k}\left( d\theta
_{i}^{2}+\sin ^{2}(\theta _{i})d\phi _{i}^{2}\right)  \label{EH d-dim}
\end{eqnarray}
where 
\begin{equation}
g(r)=1\mp \frac{r^{2}}{\ell ^{2}}~~~,~~~~~f(r)=1-\left( \frac{a}{r}\right)
^{d}  \label{d-dimmetricfns}
\end{equation}
and the cosmological constant $\Lambda =\pm d(d-1)/(2\ell ^{2})$. We shall
discuss their derivation and properties more fully in a subsequent paper %
\cite{rickrobblong}.

The EH soliton forms a new one parameter family of solutions (indexed by $p$%
) to the 5d Einstein equations (and hence low-energy string theory) with
negative cosmological constant that asymptotically approach AdS$_{5}/\Gamma $%
. These solutions are perturbatively stable and of lower energy than AdS$%
_{5}/\Gamma $. It is natural to consider the extent to which the EH soliton
has properties similar to that of the AdS soliton. The latter metric has
been shown under certain conditions to be unique lowest mass solution for
all spacetimes in its asymptotic class \cite{GSW}. It also satisfies
holographic causality \cite{PSW} and can undergo phase transitions to AdS
black holes with Ricci-flat horizons \cite{SSW}. Which of these properties
are shared by the EH soliton remains an interesting subject for future work.

\bigskip

We would like to acknowledge helpful discussions with R. Myers, K.\
Schleich, D. Witt, and E. Woolgar. This work was supported in part by the
Natural Sciences and Engineering Research Council of Canada.

\end{document}